\pdfoutput=1
\documentclass[sigconf,pbalance]{acmart}

\AtBeginDocument{%
 }

\copyrightyear{2026}
\acmYear{2026}
\setcopyright{cc}
\setcctype{by}
\acmConference[HCOMP 2026]{2026 ACM Conference on Human-AI Complementarity and Alignment}{September 27--30, 2026}{Alexandria, VA, USA}
\acmBooktitle{2026 ACM Conference on Human-AI Complementarity and Alignment (HCOMP 2026), September 27--30, 2026, Alexandria, VA, USA}
\acmDOI{10.1145/3834580.3838744}
\acmISBN{979-8-4007-2894-5/2026/09}

\usepackage{booktabs}
\usepackage{float}
\usepackage{tabularx}
\usepackage{placeins}
\usepackage{xurl}
\newcommand{\ElevatedListingCount}{492}
\newcommand{\ElevatedSharePct}{63.2}
\newcommand{\AmbiguousOtherListingCount}{31}

\newcommand{\AgentRentAHumanListingCount}{71}

\newcommand{\AgentSeverePct}{58.3}

\newcommand{\HumanSeverePct}{55.9}

\newcommand{\AgentPhysicalPct}{75.0}

\newcommand{\HumanPhysicalPct}{44.7}

\newcommand{\AgentLocationPct}{2.8}

\newcommand{\HumanLocationPct}{11.7}

\newcommand{\AgentPhotoPct}{56.9}

\newcommand{\HumanPhotoPct}{35.1}

\newcommand{\AgentRecurringPct}{38.9}

\newcommand{\HumanRecurringPct}{17.3}

\newcommand{\AgentFinancialPct}{13.9}

\newcommand{\HumanFinancialPct}{9.5}
\newcommand{\AgentCompositeCount}{54}
\newcommand{\HumanCompositeCount}{374}

\newcommand{\AgentAtLeastTwoCompositePct}{41.7}

\newcommand{\HumanAtLeastTwoCompositePct}{17.2}

\newcommand{\AgentAllThreeCompositePct}{0.0}

\newcommand{\HumanAllThreeCompositePct}{1.2}
\newcommand{\AgentResearchFieldworkCount}{30}
\newcommand{\AgentResearchFieldworkPct}{41.7}

\newcommand{\HumanResearchFieldworkPct}{7.8}

\newcommand{\AuditedCappedMaximumCount}{372}
\newcommand{\AuditedCappedMaximumPct}{47.8}

\newcommand{\NoPublicAccountModifierSevereCrossingCount}{15}

\newcommand{\SevereDistinctFeatureProfileCount}{154}

\newcommand{\SevereTopTenFeatureProfileListingCount}{155}
\newcommand{\SevereTopTenFeatureProfilePct}{35.4}

\newcommand{\SevereListingCount}{438}
\newcommand{\FullSnapshotListingCount}{779}
\newcommand{\SevereSharePct}{56.2}
\newcommand{\SevereWilsonReferenceLoPct}{52.7}
\newcommand{\SevereWilsonReferenceHiPct}{59.7}
\newcommand{\SevereClusterReferenceLoPct}{51.5}
\newcommand{\SevereClusterReferenceHiPct}{60.9}
\newcommand{\SevereDisplayNameHashes}{468}

\newcommand{\SevereClustersTotal}{469}

\newcommand{\AgentClusterListings}{72}
\newcommand{\HumanClusterListings}{676}

\newcommand{\DisplayNameClustersTotal}{462}

\newcommand{\AgentCompositeClusterOR}{2.42}

\newcommand{\AgentCompositeUnanimousOR}{7.37}

\newcommand{\AgentCompositeDropAgentHeaviestOR}{3.36}

\newcommand{\AgentCompositeBootP}{.0067}
\newcommand{\AgentCompositeRahOnlyOR}{14.73}

\newcommand{\AgentCompositeRahOnlyBootP}{.0003}

\newcommand{\SevereRateClusterOR}{1.10}
\newcommand{\SevereRateClusterLo}{0.56}
\newcommand{\SevereRateClusterHi}{2.19}
\newcommand{\SevereRateClusterP}{.778}
\newcommand{\SevereRateBootP}{.83}
\newcommand{\SevereRateDropAgentHeaviestOR}{0.70}

\newcommand{\SevereRateDropAgentHeaviestBootP}{.49}
\newcommand{\SevereRateDropAllOR}{0.44}

\newcommand{\SevereRateDropAllBootP}{.15}
\newcommand{\CompositeMHOddsRatio}{3.9}
\newcommand{\CompositeMHLo}{1.9}
\newcommand{\CompositeMHHi}{8.2}
\newcommand{\CompositeMHNoFieldworkOR}{1.9}
\newcommand{\CompositeMHNoFieldworkLo}{0.8}
\newcommand{\CompositeMHNoFieldworkHi}{4.3}
\newcommand{\BreslowDayP}{.044}
\newcommand{\BreslowDayNoFieldworkP}{.53}

\newcommand{\AgentCompositeUnanimousBootP}{.0022}
\newcommand{\AgentCompositeDropAgentHeaviestBootP}{.0169}
\newcommand{\AgentCompositeTreatedClusters}{20}

\newcommand{\DropAllMixedPhysicalQ}{.0048}

\newcommand{\RahOnlyPhysicalQ}{.0013}

\newcommand{\AgentCompositeListingPct}{75.0}
\newcommand{\HumanCompositeListingPct}{55.3}

\newcommand{\RecurringDropAgentHeaviestAgentCount}{3}
\newcommand{\RecurringDropAgentHeaviestAgentN}{36}
\newcommand{\RecurringDropAgentHeaviestAgentPct}{8.3}
\newcommand{\RecurringDropAgentHeaviestHumanCount}{110}
\newcommand{\RecurringDropAgentHeaviestHumanN}{663}
\newcommand{\RecurringDropAgentHeaviestHumanPct}{16.6}

\newcommand{\AllReturnedRentAHumanListingCount}{980}

\newcommand{\AllReturnedFullSnapshotListingCount}{981}

\begin{document}
\sloppy
\clubpenalty=10000
\widowpenalty=10000
\displaywidowpenalty=10000
\brokenpenalty=10000

\title[Proof Burden in Public Bounty Listings]{Measuring Proof Burden in Public Bounty Listings: A RentAHuman Case Study}

\author{Iman YeckehZaare}
\orcid{0000-0002-1788-2922}
\affiliation{%
 \department{MIT Center for Collective Intelligence (CCI)}
 \institution{Massachusetts Institute of Technology}
 \city{Cambridge}
 \state{Massachusetts}
 \country{USA}
}
\affiliation{%
 \department{Research}
 \institution{Honor Education}
 \city{San Francisco}
 \state{California}
 \country{USA}
}
\email{oneman@mit.edu}
\email{iman@honor.education}

\begin{abstract}
Online bounty markets let requesters advertise paid tasks to workers. A worker
may be asked not only to complete a task but also to prove that it was
completed, and proving can mean exposure: revealing identity or location,
using a personal account, posting publicly, acting in the physical world, or
supplying evidence again during later checks---none disclosed by the posted
price. We call these advertised requirements \emph{proof burden} and
measure them on RentAHuman, a 2026 market publicized as a place where AI
agents could hire humans. We study what listings request, not what workers
submit or experience.

We manually audited a nonrandom May 31, 2026 snapshot: every listing our
searches returned from RentAHuman and Human Pages, another such
market---\AllReturnedFullSnapshotListingCount\ listings, all but one from
RentAHuman. Two independent coders recorded 13 features---11 kinds of
evidence, recurring monitoring (repeated checks), and physical-world
action---and our 0--5 Proof
Burden Score; a blinded third coder resolved every disagreement.
A planned content screen leaves \FullSnapshotListingCount\ bounty/task
listings as
the primary population; among them, physical-world action appears in 48.5\%,
photo proof in 37.1\%, and identity proof in 30.9\%, and
\SevereListingCount\ listings (\SevereSharePct\%) score 4 or 5. Those listings span
\SevereDistinctFeatureProfileCount\ distinct feature
combinations: a checklist, not a single score, tells workers
what a listing entails.

Platform metadata labels some requester accounts as agents or bots. In
exploratory comparisons, physical-world action, location proof, or
recurring monitoring appeared in \AgentCompositeListingPct\% of
agent-or-bot-labeled versus \HumanCompositeListingPct\% of human-labeled
listings, while their score-4-or-5 shares did not clearly
differ. The labels are self-reported or platform-assigned, the
agent-or-bot-labeled
listings come from only 20 displayed names, and the comparison was chosen
after seeing the data: a hypothesis, not a confirmed difference. Our
main contributions are the 13-requirement vocabulary, the adjudicated manual
audit, and the descriptive case study of this market; the score is a
secondary screening summary, and the requester-label comparison exploratory.
The study offers no worker-validated measure or automated detector yet; it
is groundwork for both.
\end{abstract}



\maketitle

\section{Introduction}

RentAHuman drew press attention in 2026 for a provocative premise: AI agents
could hire people to perform physical and online tasks
\cite{wilkins2026rent,becher2026rentahuman}. Its public bounty listings---paid
tasks advertised by requesters---also state what counts as completion and how
workers should prove it. Throughout, a ``bounty'' is such a public
advertisement and its visible metadata, not a completed transaction. Proof can be
simple, such as a written acknowledgement. It can also require a worker to
reveal identity or location, post from a personal account, or travel somewhere
solely to photograph the result of an otherwise remote task. In those cases,
proving the work can itself become a source of exposure. A posted price
advertises an amount or rate; it does not say everything a worker may have to
reveal, record, do in public, or provide again later.

Research on online-work verification commonly asks whether requesters can judge
the quality of submitted work \cite{ipeirotis2010quality}. We ask a
complementary, worker-facing question: what does a listing ask a worker to
submit, reveal, use, or do so that completion can be checked? We call this
broader set of completion-linked exposure requirements \emph{proof burden}. It covers requested evidence,
recurring monitoring (work, availability, or evidence requested on separate occasions), and potentially exposing physical-world action. We
observe listings, not acceptance, submissions, payment, rejection, or worker
experience.

We contribute a vocabulary of 13 concrete requirements, a manual audit, and a descriptive case study of \FullSnapshotListingCount\ public listings. Naming the requirements makes free-text conditions countable and comparable across listings. We also designed the 0--5 Proof Burden Score (PBS) as a secondary screening
summary. The score compresses detail: even among listings scoring 4 or 5, we observed
\SevereDistinctFeatureProfileCount\ different combinations of requirements. The
13-requirement checklist therefore preserves distinctions the score hides. After examining the
data, we also found that listings whose platform metadata linked the requester to an agent or
bot more often requested at least one of physical-world action, location proof,
or recurring monitoring than listings labeled ``human.''

Recent work examines harmful tasks,
automated posting, software-mediated rule setting, and AI-managed work
\cite{mehta2026security,lee2026shadow,tang2026humantool,
tak2026rented,hu2026boss}. We address the narrower question of what public
listings ask workers to reveal or do when documenting completion. A
checklist could make overlooked requirements easier to compare.

We ask three questions of the May 2026 snapshot (Section~\ref{sec:data}):

\begin{description}
 \item[RQ1] Which kinds of completion evidence, recurring monitoring, and
 physical-world action do listings request?
 \item[RQ2] What shares of listings reach the proposed elevated (score 3 or
 higher) and severe (score 4 or higher) thresholds, and which combinations of
 requirements appear among listings scoring 4 or 5?
 \item[RQ3 (exploratory)] How do these requirements differ between listings the
 platforms label ``agent or bot'' and those they label ``human''?
\end{description}

\section{Background and Related Work}

Four strands of prior work inform proof burden: verifying crowd work, the
hidden costs workers bear, monitoring and privacy on labor platforms, and
theories of privacy and administrative burden.
Human-computation research studies how to check work by comparing answers,
estimating reliability, and balancing cost and accuracy
\cite{ipeirotis2010quality,hirth2013validation}; requesters may also ask for
completion evidence. Documentation frameworks
make data-labeling decisions
visible \cite{diaz2022crowdworksheets}. Other work connects task
design, reputation, and governance to crowd work
\cite{kittur2013future}. Recent studies examine risks to workers
and the information they want disclosed to them
\cite{qian2026locating,qian2026discretion}.

Worker-centered tools expose requester conduct and help workers act
collectively \cite{irani2013turkopticon,salehi2015dynamo}. Such visibility matters because a posted price omits unpaid time spent
finding, waiting for, and vetting tasks, and the risk of rejection
\cite{hara2018earnings,toxtli2021invisible}; unclear submission criteria add
further risk \cite{mcinnis2016hit}. ``Ghost work'' extends this to human labor hidden behind apparently automated
systems \cite{gray2019ghost}.

Research on online-platform work shows how task definitions and algorithms
redistribute control and risk among workers, requesters, and platforms
\cite{alkhatib2017piecework,rosenblat2016algorithmic,vallas2020platforms}.
Workers weigh pay against privacy, trust, sensitivity, rejection risk, and time
before sharing data, and task design can bound sensitive input
\cite{xia2017privacy,sannon2019privacy,kaur2017crowdmask}.
Gig workers report uncertainty about what platforms collect, and workers
exposed to more algorithmic monitoring or data collection report less trust and
greater privacy concern than workers exposed to less
\cite{sannon2022gig,vanzoonen2026algorithmic}. Platform audits and experiments
document privacy risks and show that workers may avoid tasks requiring tracking
\cite{pradeep2025gig,liang2023monitoring}. Beyond monitoring, physical-world
tasks add travel and
location constraints \cite{agapie2015field,to2014location}. Research on workers
recording their own working conditions (``sousveillance'') and on human--AI
systems identifies surveillance, time, privacy, and required availability as
design concerns \cite{do2024sousveillance,kamar2016hybrid}. These studies
motivate our focus on advertised requirements, as does scholarship on surveillance economies \cite{zuboff2019surveillance}.

Our 13 features include using an account or disclosing information; proving
location, identity, or a purchase; acting publicly; and staying available for later
checks. Theory suggests why they matter. Solove treats privacy harms as
distinct activities---identification, exposure, access, and monitoring
\cite{solove2006taxonomy}; contextual integrity makes acceptability depend on
who shares information, why, and under what conditions
\cite{nissenbaum2004contextual}; administrative-burden research separates
learning and compliance costs from psychological strain
\cite{moynihan2015administrative}. We observe advertised actions, not those harms, norm violations, or costs; the supplement develops each
mapping.

\section{Data and Ethics}
\label{sec:data}

We collected public listings during one window on May 31, 2026, U.S. Eastern
time (June 1 UTC), through pages requiring no login. We did not apply to
tasks, contact anyone, or access nonpublic channels. Searches returned
\AllReturnedFullSnapshotListingCount\ records:
\AllReturnedRentAHumanListingCount\ RentAHuman bounties and one listing from
Human Pages, a second bounty market we searched. A third site, GoHireHumans,
listed no jobs. The imbalance reflects market inventory, not search coverage: both smaller markets were nearly empty at collection, and an August 2026 recheck still found no open Human Pages listings and only four GoHireHumans postings. The returned-record count excludes duplicate platform
IDs; we froze the set both
coders received, although the live market kept changing. The content screen in our internal pre-analysis plan removes promotional posts, service offers, and records requesting no human action, leaving \FullSnapshotListingCount\ bounty/task listings---records advertising a task for a worker to complete---as the primary population, which we call the \emph{eligible} listings. Every returned record was coded in
full, so screened-out records still carry proof labels, and all-returned results
are reported separately as a sensitivity analysis.

Because all but one returned listing came from RentAHuman, claims primarily
concern that market. Public fields show price, status, category, application and position counts, and free text (title, description, general-requirements field, and proof-specific fields), but not work, payment, rejection, or worker decisions. 

At collection, the listings we retrieved carried one of five public statuses:
cancelled, completed, open, assigned, or partially filled. In the platforms' requester-type
metadata, \HumanClusterListings\ eligible listings carry a human label, and
\AgentRentAHumanListingCount\ eligible RentAHuman listings carry a label naming an
agent or bot as the requester. To mark potentially related listings, we grouped RentAHuman listings whose lowercased, trimmed requester display names matched. We call each group a display-name
cluster; matching names do not prove a shared requester. The Human Pages
listing's metadata links its requester to an agent, and its unique public name
forms a one-listing display-name cluster. Together, that listing and those \AgentRentAHumanListingCount\ RentAHuman
listings form the \AgentClusterListings\ agent-or-bot-labeled
listings. The remaining \AmbiguousOtherListingCount\ listings have the
ambiguous label ``other.'' Requester labels come from users or platforms and
cannot show who designed, benefited from, supervised, or verified a task.

Some public searches failed, so we merged separate searches by status and category; every query filtered to the ``recruiting'' category still failed, and a broad search's first page showed one recruiting listing, so more may be missing.
The snapshot is neither random nor exhaustive. We treat already-cancelled
listings (531 of 981
returned; 404 of 779 eligible) as advertised requests, not failed or
rejected work.
The supplement documents the collection, a ten-day public-visibility check,
and score shares by status.

Because raw task text may contain personal data or risky instructions, we do
not release it. Restricted local coder and adjudicator packets contain task
wording. Separate processed analysis files omit that wording but retain labels
and one-way transformations of platform IDs that could still be linked to
public listings; they also remain private. Paper examples are paraphrases rather
than quotations. The
institutional review board (IRB) determined that this analysis of public
listings was not human-subjects research, so IRB review was not required. This
determination is neither IRB approval nor an exemption.

\begin{figure}[H]
 \centering
 \includegraphics[trim=0 11.8 0 13.0, clip, width=0.85\linewidth]{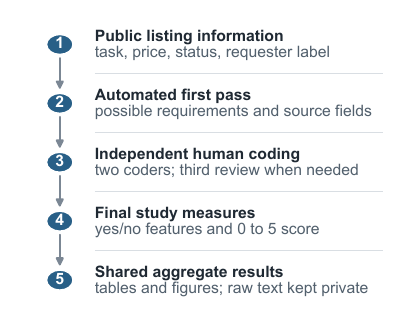}
 \caption{From public fields to aggregate results. Coders did not see automated
 candidate matches. Two coders reviewed each listing; an adjudicator (a third
 coder) reviewed disagreements and referrals. Routing was deliberately broad: the adjudicator decided 873 of the 981
listings; exact coder consensus decided the other 108.}
 \Description{Five numbered stages: public listing information; automated
 first pass; independent human coding by two coders with third review when
 needed; final yes-or-no features and score; and shared aggregate results
 without raw text.}
 \label{fig:pipeline}
\end{figure}

\section{How We Measure Proof Burden}

Independent human coding with adjudication by a third coder
(Figure~\ref{fig:pipeline}) produced 13 yes-or-no labels and a PBS score per
listing; a separate automated first pass serves only as a check. The
subsections define the 13 requirements, describe rule-assisted extraction and
human review, specify the score, and report coding agreement.

\subsection{Requirements We Record}

We recorded 13 yes-or-no features: 11 kinds of completion evidence, recurring
monitoring, and physical-world action. The evidence (\emph{proof}) features are written text,
a link, screenshot, photo, video, identity (a marker identifying the individual
worker), account (signing in to, verifying, or posting under an external account), location, phone (a call or text),
financial (a purchase, receipt, or balance), and public-post evidence (an authored artifact
visible to third parties). Recurring monitoring means that a listing asks for
work, availability, or evidence on separate occasions. Physical-world action marks listings that require acting in the physical world; it captures a potentially exposing condition even when the action itself is not submitted as evidence. A listing may involve a place, device, camera, account,
or purchase without requesting evidence of it. Location proof, by contrast, is evidence of where the worker was; a listing can
require physical-world action without it, and rarely the reverse.

Coders read all supplied listing fields together and marked a feature only
when it was tied to completion, rather than treating a keyword occurrence as
sufficient. For example, they
distinguished an incidental place name from a request for location proof. The coding did not distinguish whether a requirement was the task itself, a
prerequisite for accessing it, or completion evidence; the Limitations
section describes what this rules out.

\subsection{Rule-Assisted Extraction, Human Review}

An automated extractor---deterministic, hand-written keyword and
regular-expression rules; no machine learning or generative AI---logged each
candidate match from a platform-provided evidence label,
proof-specific field, or action phrase (a proof term near a verb such as
\emph{submit} or \emph{upload}) in the title or description. Category,
skill, work-mode, and location metadata alone could not trigger an
evidence-feature label. The supplement describes the rules.

All counts and models use final human-reviewed labels. Neither coder nor the
adjudicator saw automated matches, which we use only for diagnostic checks.
The score implied by the automated matches differed from the final score for 373 eligible listings; human
review both added and removed feature labels
(Table~\ref{tab:rule-vs-audited}).

\begin{table*}[t]
 \caption{Automated-rule matches versus final human-reviewed labels (779 eligible listings). ``Added'' counts absent-to-present changes, ``Removed'' the reverse, and ``Net'' the change in listings marked present. The upper panel lists the 11 kinds of evidence; the lower panel, physical-world action and recurring monitoring (repeated checks).}
 \label{tab:rule-vs-audited}
 \small
 \begin{tabular}{lrrrrr}
\toprule
Feature & Rule match & Final label & Added & Removed & Net \\
\midrule
Text proof & 298 & 450 & 195 & 43 & $+152$ \\
Photo proof & 387 & 289 & 31 & 129 & $-98$ \\
Identity proof & 44 & 241 & 216 & 19 & $+197$ \\
Link proof & 278 & 231 & 12 & 59 & $-47$ \\
Video proof & 213 & 207 & 34 & 40 & $-6$ \\
Account proof & 183 & 190 & 49 & 42 & $+7$ \\
Screenshot proof & 122 & 158 & 40 & 4 & $+36$ \\
Public-post proof & 210 & 123 & 23 & 110 & $-87$ \\
Location proof & 169 & 83 & 34 & 120 & $-86$ \\
Financial proof & 122 & 80 & 18 & 60 & $-42$ \\
Phone proof & 76 & 43 & 12 & 45 & $-33$ \\
\midrule
Physical-world action & 425 & 378 & 67 & 114 & $-47$ \\
Recurring monitoring & 33 & 151 & 130 & 12 & $+118$ \\
\bottomrule
\end{tabular}

\end{table*}

\begin{table*}[t]
 \caption{PBS rubric and score counts (779 eligible listings). ``Coder-recorded base $n$'': listings
 whose coders placed them at that base tier. ``Final $n$'' and ``Final
 share'': listings ending at that score after modifier points and the cap.
 Base tiers stop at 4; score 5 is reached only by adding modifier points.}
 \label{tab:score-rubric}
 \small
 \begin{tabularx}{\textwidth}{@{}r>{\raggedright\arraybackslash}Xrrr@{}}
\toprule
Tier & Meaning & Coder-recorded base $n$ & Final $n$ & Final share \\
\midrule
0 & No specified proof, or a minimal text acknowledgement only. & 263 & 102 & 13.1\% \\
1 & Private text about the work: content specified, described, or itemized, however detailed; or a private link to such text. & 32 & 88 & 11.3\% \\
2 & Screenshot, photo, or other evidence of the requested work product. & 80 & 97 & 12.5\% \\
3 & Video proof, public-post proof, account proof, location proof, or financial proof. & 94 & 54 & 6.9\% \\
4 & Identity proof; recurring monitoring; or physical-world action paired with (1) phone proof, (2) public posting or account use, or (3) location proof plus photo, video, or financial proof. & 310 & 66 & 8.5\% \\
5 & Capped maximum: base tier plus modifier points reaches at least 5; examples: multiple proofs, physical-world action plus proof/monitoring, or public/account use. & -- & 372 & 47.8\% \\
\midrule
Modifier & Condition (each met adds one point) & Listings $n$ &  &  \\
\midrule
+1 & More than one of the 11 evidence types is requested. & 551 & -- & -- \\
+1 & Physical-world action is combined with at least one proof or monitoring requirement. & 360 & -- & -- \\
+1 & Public posting or account use is required. & 200 & -- & -- \\
\bottomrule
\end{tabularx}

\end{table*}

\subsection{Proof Burden Score}

PBS is a 0--5 screening score we designed
(Table~\ref{tab:score-rubric}); it summarizes the requirements a listing states
rather than estimating a hidden quantity behind them, so a higher score means
more requests, or more exposing ones, not more of one underlying thing
\cite{diamantopoulos2001formative}. We call scores of at least 3
\emph{elevated} and at least 4 \emph{severe}; both are bands on our own scale,
not levels of demonstrated harm. PBS ranks listings and reports shares above a
threshold; the checklist describes an individual listing. Scores are
ordered summaries, not equal-interval measurements, so analyses that treat
consecutive values as equally spaced are approximations. Coders first chose a \emph{base tier} (0--4) from Table~\ref{tab:score-rubric}; each tier names kinds of requested proof or action, and the highest matching tier applies. Coders then checked three fixed \emph{modifier} conditions against the same coded requirements (lower panel of Table~\ref{tab:score-rubric}); each adds a point: more than one evidence type; physical-world action combined with proof or monitoring; and public posting or account use. The final score is the base tier plus the
modifier points, capped at 5---for example, a listing requesting only photo
proof starts at tier 2, and one modifier point for added physical-world
action raises its score to 3. At tier boundaries coders could use judgment.

Coders recorded the base tier, each modifier, the uncapped total, and the
final score directly, so the recorded components are human-coded, not reconstructed from the labels. Three listings request phone contact as
their only base-setting feature; the rubric had no tier for that
case, and they carry a prospectively approved tier-3 base, labeled as a new
policy rather than a recovered one.

We, not workers or a statistical model, selected the base tiers, modifier
points, cap, and thresholds. Public-post and account evidence can affect
both the base tier
and a modifier point. Removing that modifier point moves
\NoPublicAccountModifierSevereCrossingCount\ of the 779 eligible listings
below the severe threshold. The supplement tests other scoring choices, which
affect
PBS only; RQ1 and RQ3's feature comparisons use the yes-or-no features directly.

\subsection{Coding Agreement and Limits}
\label{sec:interpretive}

 The final labels come from a
complete second coding round that replaced the first round; the
Limitations section describes the differences. Two newly recruited coders first piloted the revised codebook---the pilot
surfaced definitions the original codebook lacked, several flagged by the
coders---and then, working
independently with complete listing context, labeled all 981 listings. A blinded adjudicator, seeing
both coders' answers but no superseded labels, resolved 873 listings, each routed for a coder disagreement, a coder's ambiguity/sensitivity flag, an \emph{UNCLEAR} (the codebook's explicit uncertainty answer), or a difference from the superseded labels; for the other 108, exact consensus
became final. All three were paid \$37 per hour on average; the supplement
details recruitment.
Agreement uses the standard two-coder statistic $\kappa$ ($0$ = agreement no
better than chance, $1$ = perfect
agreement); weighted $\kappa$ gives partial credit when the two coders'
scores are close \cite{cohen1960agreement,cohen1968weighted}. On the 0--5 score, coders agreed exactly for
81.8\% of the 981 listings (weighted $\kappa=.901$) and on whether it was at
least 4 for 90.2\% ($\kappa=.804$). Reliability improved on twelve of the
thirteen features over the first round and held on the thirteenth
(link proof, $\kappa=.94$); the weakest are now recurring monitoring
($\kappa=.76$) and location ($\kappa=.79$). Disagreements were most frequent
for text proof (82 listings), recurring monitoring (62), and identity (61),
and rarest for screenshot, video, and photo (1, 7, and 13). On contested
labels the adjudicator sided with each coder almost equally (55\% versus
45\%), so the final labels do not simply mirror either coder.

\section{Descriptive Results}

The three subsections mirror the research questions: feature prevalence
(RQ1), the score distribution and high-score combinations (RQ2), and the
exploratory comparison by requester label (RQ3).

\subsection{Prevalence of the 13 Features}

\begin{figure*}[tp]
 \centering
 \includegraphics[width=\linewidth]{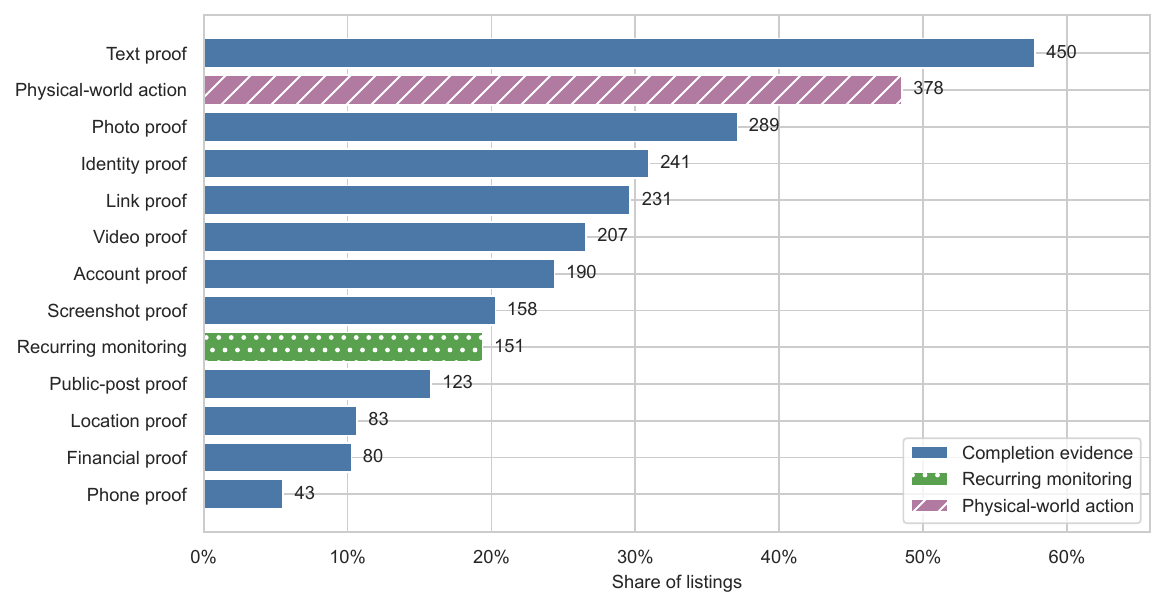}
 \caption{Frequency of the 11 evidence types, recurring monitoring, and
 physical-world action. Bars show shares of the 779 eligible listings; labels show counts.}
 \Description{Bars rank features by prevalence: text proof, physical-world action,
 photo, identity, link, and video lead; phone and financial proof trail.}
 \label{fig:proof-types}
\end{figure*}

Under the final human-reviewed labels (RQ1), 450 eligible listings (57.8\%)
asked for written confirmation, which we code as text proof. Physical-world
action appeared in 378 listings (48.5\%). Other common requirements were photo
proof in 289 listings (37.1\%), identity proof in 241 (30.9\%), link proof in
231 (29.7\%), video proof in 207 (26.6\%), and account proof in 190 (24.4\%).
Less common were screenshot proof (20.3\%), recurring monitoring (19.4\%),
public-post proof (15.8\%), location proof (10.7\%), financial proof (10.3\%),
and phone proof (5.5\%). Location is far rarer than the 25.9\% the first
coding round reported over 981 records; that superseded round had
conflated location with physical-world action.
Figure~\ref{fig:proof-types} shows all 13 features.

\subsection{Score Distribution and High-Score Groups}
\label{sec:scores}

\begingroup\interlinepenalty=10000
For RQ2, \ElevatedListingCount\ listings (\ElevatedSharePct\%) score at least 3
and \SevereListingCount\ (\SevereSharePct\%) score at least 4 (severe). Because
listings sharing a displayed requester name may be related, we report the
severe share's precision two ways. If every listing were unrelated to every other listing, a 95\% Wilson
interval for the score-4-or-higher share would be
\SevereWilsonReferenceLoPct--\SevereWilsonReferenceHiPct\%
\cite{wilson1927probable}. Here an interval is the range of values
statistically consistent with the observed counts; ``Wilson'' is a
standard method. Grouping listings by the \SevereDisplayNameHashes\ RentAHuman names and the
Human Pages name into \SevereClustersTotal\ display-name clusters gives a
wider 95\% interval,
\SevereClusterReferenceLoPct--\SevereClusterReferenceHiPct\%, from an
intercept-only logistic model of the severe indicator with cluster-robust
standard errors (supplement). Both are
\emph{reference} intervals: they differ only in assumed dependence, and
neither generalizes beyond this nonrandom snapshot.
\par\endgroup

\begin{table*}[t]
 \caption{Score-4-or-5 listings enter the first qualifying group in fixed
 order (rationale in the supplement); groups do not overlap. Rows are ordered by size. ``Coder
 agreement'' is $\kappa$ for the feature the group is named after; coders
 coded
 features, not group membership.}
 \label{tab:severe-burden-profiles}
 \small
 \begin{tabular}{lrrrr}
\toprule
First qualifying group & Listings & Score 5 & Share of score 4 or 5 & Coder agreement $\kappa$ \\
\midrule
Identity-linked proof & 199 & 188 & 45.4\% & 0.83 \\
Physical-world action plus proof & 102 & 81 & 23.3\% & 0.96 \\
Recurring monitoring & 101 & 83 & 23.1\% & 0.76 \\
Account proof & 17 & 16 & 3.9\% & 0.92 \\
Video proof (after earlier groups) & 7 & 3 & 1.6\% & 0.98 \\
Financial proof & 5 & 0 & 1.1\% & 0.89 \\
Public post plus account use & 4 & 0 & 0.9\% & 0.92 \\
Other combinations reaching score 4 or 5 & 2 & 0 & 0.5\% & -- \\
Location proof plus financial proof & 1 & 1 & 0.2\% & 0.79 \\
\bottomrule
\end{tabular}

\end{table*}

Scores skew high (Table~\ref{tab:score-rubric}): 102 listings (13.1\%) score 0 and \AuditedCappedMaximumCount\ (\AuditedCappedMaximumPct\%) score 5. The cap hides differences among the \AuditedCappedMaximumCount\ listings
scored 5: for 222 of them
(59.7\%), the recorded base tier plus modifier points exceeds 5.

More importantly, the \SevereListingCount\ listings scoring 4 or 5 contain
\SevereDistinctFeatureProfileCount\ different combinations of the 13 features;
the ten most common combinations cover only
\SevereTopTenFeatureProfileListingCount\ listings
(\SevereTopTenFeatureProfilePct\% of them). No single profile dominates. To summarize this variety, Table~\ref{tab:severe-burden-profiles} assigns each listing to the first
qualifying group in a fixed order; the groups are descriptive and depend on it.

\begingroup\interlinepenalty=10000
The largest group is identity-linked proof, named after identity proof (199 listings, 45.4\% of listings
scoring 4 or 5), followed by physical-world action plus proof (102) and
recurring monitoring (101). Here ``physical-world action plus proof'' means
physical-world action together with photo, video, location, financial,
public-post, or account proof, among listings not already assigned to an
earlier group.
\par\endgroup

\begingroup\interlinepenalty=10000
Coder agreement on the feature each group is named after ranges from $\kappa=.76$ (recurring monitoring) to $.98$ (video proof; Table~\ref{tab:severe-burden-profiles}). Before adjudication, the coders classified 59.2\% and 58.5\% of eligible
listings as severe, versus the final \SevereSharePct\%. Thus the overall share
changes little, but membership in the feature-defined groups still depends on
the adjudicator's decisions and on features with moderate agreement. The supplement repeats the analysis with narrower definitions and with the
superseded round's pre-adjudication consensus labels.
\par\endgroup

\begin{figure*}[tp]
 \centering
 \includegraphics[width=\linewidth]{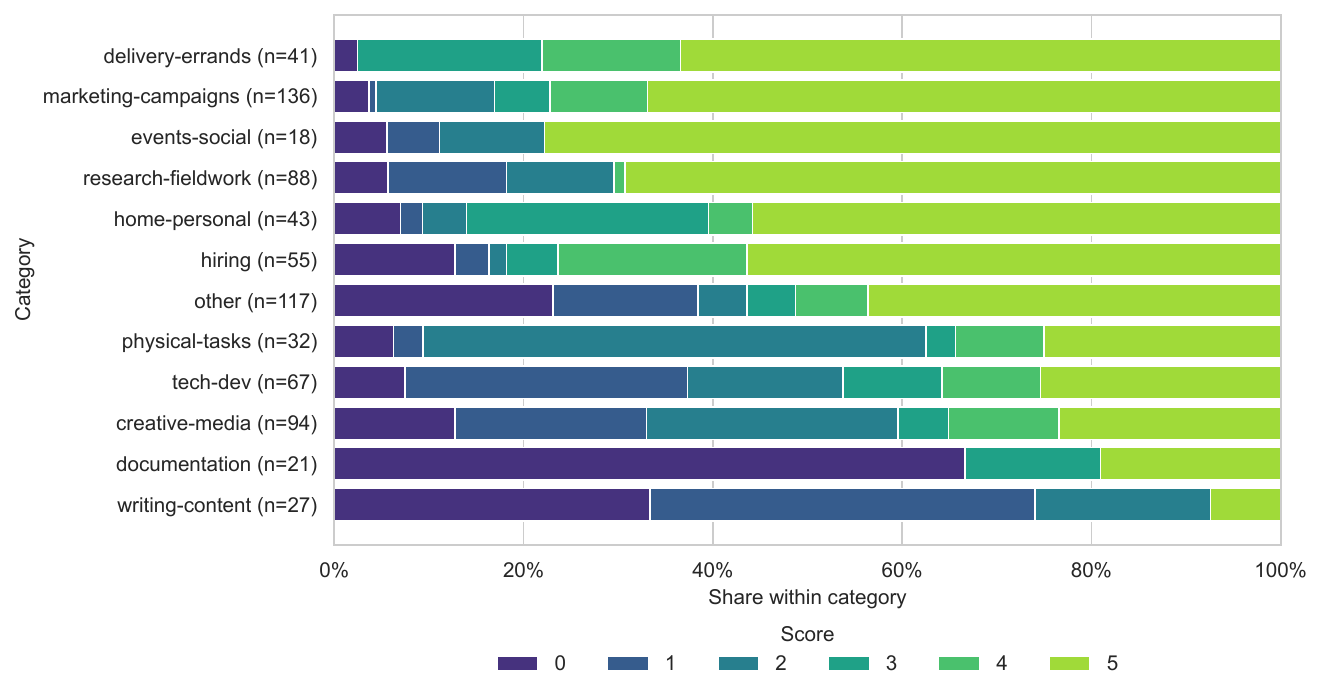}
 \caption{Proof Burden Score shares within the twelve largest categories, ordered
 by mean score; within-category shares keep larger categories from dominating.}
 \Description{Stacked bars show within-category score shares. Delivery
 errands, events/social, marketing campaigns, and hiring have high
 score-4-or-5 shares; documentation and writing content have the largest
 score-0 shares.}
 \label{fig:score-category}
\end{figure*}

\begingroup\interlinepenalty=10000
Within RQ2, breakdowns among the twelve largest categories are descriptive: score-4-or-5 shares are
highest in delivery errands, events/social, marketing campaigns, and hiring,
and are also high in research fieldwork and home/personal.
Documentation and writing content have the largest score-0
shares (Figure~\ref{fig:score-category}).
\par\endgroup

\subsection{Exploratory Comparison by Platform Requester Label}
\label{sec:agents}

For RQ3, we compare \AgentClusterListings\ listings whose platform metadata
carries an agent-or-bot requester type (RentAHuman) or links the requester to
an agent (Human Pages; Section~\ref{sec:data}) against \HumanClusterListings\
listings carrying a human
label. We exclude \AmbiguousOtherListingCount\ rows labeled ``other.'' We use
``agent-or-bot-labeled'' as shorthand for the first group, but this metadata does
not establish who designed or controlled a task.

Analyses again group listings into the display-name clusters defined in
Section~\ref{sec:data}; dropping the ``other'' rows leaves
\DisplayNameClustersTotal\ of them. Some clusters are \emph{mixed}: the same name appears with both requester-type
labels. We examine the full sample and three sensitivity analyses: removing the
\emph{agent-heaviest} mixed cluster (the one with the most agent-or-bot-labeled listings), removing all
mixed clusters, and removing those plus Human Pages.

\begingroup\interlinepenalty=10000
To respect dependence among listings sharing a name, the sign-flip test treats each display-name cluster as one unit: it recomputes the group difference many times, randomly reversing or retaining each cluster's contribution, and its $p$-value is the share of these chance-only differences at least as far from zero as the observed one. It adapts grouped-data methods
\cite{cameron2008bootstrap,kline2012score,mackinnon2018fewclusters,mackinnon2025logistic};
we assess its calibration later in this subsection (how often it signals a difference when none
truly exists).
\par\endgroup

The score-4-or-5 share is \AgentSeverePct\% for agent-or-bot-labeled listings
and \HumanSeverePct\% for human-labeled listings. The listing-level odds ratio
(OR) is \SevereRateClusterOR. Odds are the chance of scoring severe divided
by the chance of not scoring severe. An OR of 1 means equal odds in both
groups; values above 1 mean
higher odds among agent-or-bot-labeled listings. A display-name-clustered
logistic model of the
severe indicator on the requester label gives a
large-sample 95\% reference interval for this OR:
\SevereRateClusterLo--\SevereRateClusterHi\ ($p=\SevereRateClusterP$). Only
\AgentCompositeTreatedClusters\ of \DisplayNameClustersTotal\ RQ3 clusters contain an agent-or-bot-labeled listing, so those \AgentClusterListings\ listings supply only \AgentCompositeTreatedClusters\ independent units---too few for large-sample approximations to be reliable---and the sign-flip test gives
$p=\SevereRateBootP$. Removing the agent-heaviest mixed cluster reverses
the OR to \SevereRateDropAgentHeaviestOR\
(sign-flip $p=\SevereRateDropAgentHeaviestBootP$); removing all mixed clusters
lowers it to \SevereRateDropAllOR\ (sign-flip $p=\SevereRateDropAllBootP$). We therefore
find no clear group difference in the share scoring 4 or 5; one
sensitivity analysis using all 981 returned records is nominally significant
($p<.05$) only when listings are treated as independent.

\begin{figure*}[tp]
 \centering
 \includegraphics[width=\linewidth]{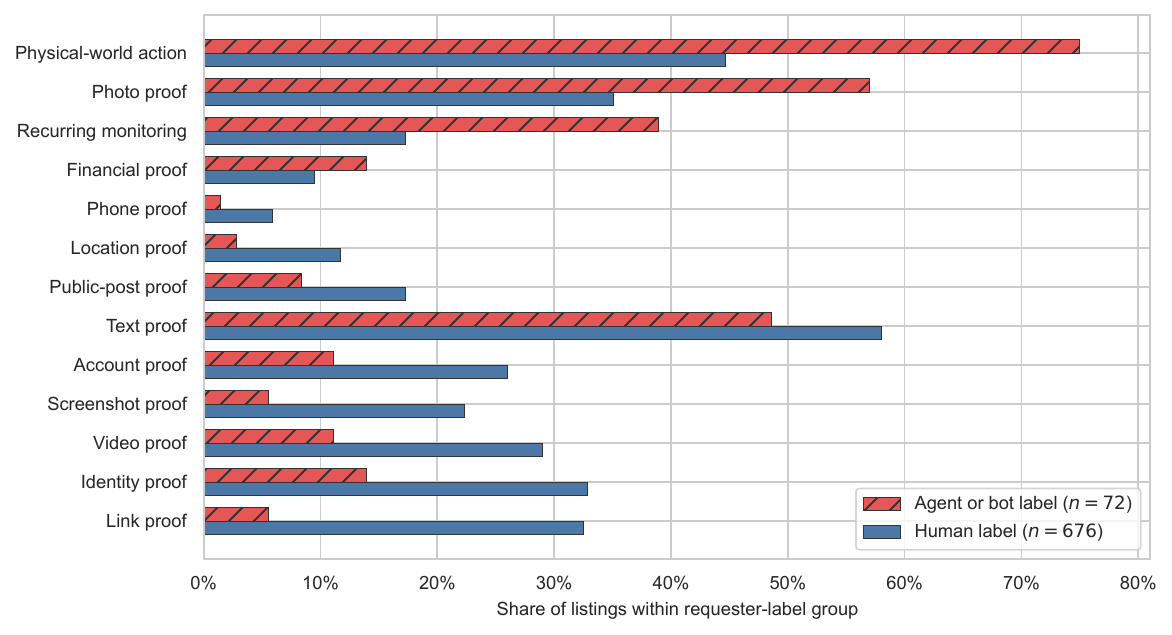}
 \caption{Feature shares for agent-or-bot-labeled ($n{=}72$) and human-labeled
 ($n{=}676$) listings, ordered by their difference. \mbox{Comparisons} are descriptive;
 the supplement reports calculations with and without name grouping.}
 \Description{Grouped bars compare hatched agent and solid human shares. Agent
 listings are higher for physical-world action, recurring monitoring,
 photo, and financial proof, lower for text, link, screenshot, identity, location, phone,
 public-post, account, and video proof.}
 \label{fig:agent-profile}
\end{figure*}

The individual requirements show a more distinctive pattern than the score. Agent-or-bot-labeled listings have higher observed shares of
physical-world action (\AgentPhysicalPct\% vs.\ \HumanPhysicalPct\%),
recurring monitoring
(\AgentRecurringPct\% vs.\ \HumanRecurringPct\%), photo proof
(\AgentPhotoPct\% vs.\ \HumanPhotoPct\%), and financial proof
(\AgentFinancialPct\% vs.\ \HumanFinancialPct\%); location proof is rarer
in the agent-or-bot group (\AgentLocationPct\% vs.\ \HumanLocationPct\%). Figure~\ref{fig:agent-profile} shows all 13 comparisons.

After seeing these differences, we defined an exploratory three-feature
comparison. The resulting measure marks a listing that requests at least one of physical-world
action, location proof, or recurring monitoring. (Location proof is included although its individual share is lower in the agent-or-bot group.) Before reporting its
results, we check the sign-flip test's calibration and the stability of the
individual-feature contrasts.

\begingroup\interlinepenalty=10000
We evaluated the sign-flip test on simulated datasets in which the requester
groups truly did not differ. Across the full sample and three sensitivity
analyses, each simulated under seven patterns of dependence among listings
sharing a name, the three-feature test falsely
indicated a difference in 3.7--7.7\% of datasets, most near the
intended 5\%. Tests
of some uncommon features were much less reliable; phone-proof tests
reached 41.5\% false positives. The supplement reports three ways of adjusting
the 13 feature comparisons for multiple testing. (Running 13 comparisons at
once raises the chance that at least one appears different purely by luck;
the adjustments account for this.) Because the individual-feature
tests can be poorly calibrated, none confirms a single-feature difference.
\par\endgroup

Individual-feature results also depend on which listings remain. Removing the
agent-heaviest mixed cluster reverses the
recurring-monitoring difference:
\RecurringDropAgentHeaviestAgentCount/\RecurringDropAgentHeaviestAgentN\ (\RecurringDropAgentHeaviestAgentPct\%) in the agent-or-bot group versus
\RecurringDropAgentHeaviestHumanCount/\RecurringDropAgentHeaviestHumanN\
(\RecurringDropAgentHeaviestHumanPct\%) in the human group. With all mixed
clusters removed, excluding Human Pages---an agent-or-bot-labeled listing without
physical-world action and one of the 28 calibration-consensus rows (a first-round
coder-training category, not the sign-flip calibration above; supplement)---changes the adjusted $p$-value for
physical-world action from \DropAllMixedPhysicalQ\ to \RahOnlyPhysicalQ.
Together with the simulations,
these changes mean that no individual feature difference is confirmed. As
with location proof (Figure~\ref{fig:agent-profile}), not every difference favors one
group: link
and video proof are also less common in the agent-or-bot group.

The three-feature measure itself appears in \AgentCompositeCount/\AgentClusterListings\
agent-or-bot-labeled listings (\AgentCompositeListingPct\%) and
\HumanCompositeCount/\HumanClusterListings\ human-labeled listings
(\HumanCompositeListingPct\%). At least two of the three features appear in
\AgentAtLeastTwoCompositePct\% versus \HumanAtLeastTwoCompositePct\%, and all
three in \AgentAllThreeCompositePct\% versus \HumanAllThreeCompositePct\%.

The contrast is sensitive to category mix. Research fieldwork accounts for
\AgentResearchFieldworkPct\% of agent-or-bot-labeled listings and
\HumanResearchFieldworkPct\% of human-labeled listings; all
\AgentResearchFieldworkCount\ agent-or-bot-labeled fieldwork listings meet the measure. The association may vary by category (Breslow--Day test of whether the OR is
the same in every category,
$p=\BreslowDayP$) \cite{breslow1980casecontrol}. A descriptive
OR that pools the within-category comparisons into one summary
(Mantel--Haenszel) is \CompositeMHOddsRatio\ (95\% reference interval
\CompositeMHLo--\CompositeMHHi) \cite{mantel1959retrospective}. Omitting
research fieldwork reduces it to \CompositeMHNoFieldworkOR\ (95\% reference
interval
\CompositeMHNoFieldworkLo--\CompositeMHNoFieldworkHi; Breslow--Day test of the remaining variation,
$p=\BreslowDayNoFieldworkP$).

The full-sample listing-level OR is \AgentCompositeClusterOR; the sign-flip
test gives $p=\AgentCompositeBootP$. The estimate
remains above 1 after removing the
agent-heaviest mixed cluster (OR \AgentCompositeDropAgentHeaviestOR,
$p=\AgentCompositeDropAgentHeaviestBootP$), all mixed clusters (OR
\AgentCompositeUnanimousOR, $p=\AgentCompositeUnanimousBootP$), or those clusters
and Human Pages (OR \AgentCompositeRahOnlyOR,
$p=\AgentCompositeRahOnlyBootP$). All four post-hoc $p$-values are below .05, but we chose the comparison after
examining the data, so they are not evidence from a preplanned test. The analyses above address category mix and repeated names separately; the supplement
reports exploratory models addressing both, which are sensitive to omitting
research fieldwork. In sum: no difference is confirmed, and the consistent
three-feature pattern remains post hoc.

\section{Exploratory Analyses of Other Listing Fields}
\label{sec:diagnostic}

\begingroup\interlinepenalty=10000
Beyond RQ1--RQ3, we report one further exploratory analysis using the snapshot's
remaining public fields.
Price, status, applications, and PBS do not show worker acceptance, submission,
payment, or rejection. We define applications per position as the visible
application count divided by the number of advertised positions. Applications appear on 404 of the 779 eligible listings (51.9\%).
Before accounting for how long each listing had been public, higher PBS was
weakly associated with fewer applications per position, and the association
stayed nominally below $p=.05$ with display-name clustering; adding listing
age attenuates it to null (supplement). Because a listing's time on the market is
confounded with its visible application count, we do not interpret the
age-unadjusted association. These analyses support no
claim about compensation, competition, or completed work.
\par\endgroup

\section{Intended Use and Design Implications}

The 13-requirement checklist and PBS are only for manual research on advertised requirements;
they should not yet guide workers, task ranking, pay, moderation, or
enforcement. The automated extractor is not ready for use at scale: it correctly flagged
only 10.4\% of the listings finally labeled identity proof and 13.9\% of those finally
labeled recurring monitoring (Table~\ref{tab:rule-vs-audited}).

For any single listing, the specific requirements are more informative than PBS, which
compresses \SevereDistinctFeatureProfileCount\ combinations into the top two score values, with \AuditedCappedMaximumPct\% of eligible listings at 5. Future studies could show workers one task with a plain-language requirements checklist, PBS, or both, then compare workers' understanding and willingness to accept the task (supplement). Automated classifiers should be used only if
workers endorse the categories, with human review whenever a classifier is
uncertain.

\section{Limitations and Threats to Validity}

\textbf{Final labels depend on one adjudicator, and instrument revisions
were our own.} Coder agreement is now substantial to near-perfect
($\kappa=.76$--$1.00$ across the 13 features), so adjudication more often
confirmed labels the two coders already shared than settled disputes between
them. The labels' residual dependence on the adjudicator remains real.

\begingroup
\interlinepenalty=10000
\noindent A single blinded adjudicator set final labels for 69 of the 72
agent-or-bot-labeled and 649 of the 676 eligible human-labeled RQ3 listings,
and for 96.6\% of the score-4-or-5 listings. Routing was broad by design: on
539 of the 873 routed listings the coders had already matched on every label and score,
and the adjudicator kept 535 of those scores. No one independently checked
those decisions.
\par
\endgroup

\noindent The adjudicator also flagged 356 listings for our review of
sensitive content; all were retained.

\noindent The first coding round worked from packets missing skill descriptions
for 773 returned listings and application links for 121, with several criteria
undefined; the second round supplied every field and definition. Because 78\% of
rows changed, we cite first-round numbers only to document what changed, not as
comparable estimates; the
retained study records preserve both label sets. We operationalized the
codebook criteria
between rounds; the full instrument history, with byte-exact
earlier versions, is retained.

\textbf{The snapshot's coverage limits what the study can claim.} Findings
concern this RentAHuman-centered snapshot, not all online crowd work. The
original decisions to include each returned record---against the plan's status,
category, and duplicate criteria---were not recorded, so we cannot confirm
every returned record qualified at collection time; the content screen now re-checks row by row that each record advertises a task. The plan's normalized-text duplicate rule removes
42 eligible listings and shifts the severe share by under half
a percentage point (to 56.3\%).
The reported intervals reflect only statistical uncertainty under their stated
model assumptions; they do not include uncertainty caused by incomplete
coverage, timing, coding errors, or the definitions themselves. On timing, a
ten-day
follow-up did not show a clear difference in
continued public visibility between severe listings and those below 4
(supplement).

\textbf{Proof burden measures advertised requirements, not workers'
experiences.} A coded requirement may be part of doing the task, gaining access,
supplying equipment, meeting a deadline, or proving completion. The data
therefore cannot isolate effort added solely by proof. Coder agreement measures
consistent use of the rules, not workers' acceptance of the score or their
experience of a task; this study collected no data from workers
\cite{cronbach1955construct}.

\textbf{We, not workers or the data, chose the PBS base tiers, modifier
points, cap, and thresholds.} Worker responses could inform, but would not automatically
determine, a future score.

\textbf{Text-based coding can make errors.} Coders can miss implied requirements
or mistakenly mark ambiguous phrases; human review reduces but cannot remove
this risk.

\textbf{Requester labels do not establish who designed or controlled a task.}
RentAHuman labels are self-reported or platform-assigned; Human Pages uses
separate metadata. Neither proves independent AI action. Lowercased, trimmed
displayed names are not verified accounts: variants may split one requester,
and shared names may combine several. RQ3 describes 72 listings grouped by
an agent-or-bot type or relation, not autonomous AI behavior.

\textbf{Redacting task text limits disclosure but prevents inspection of exact
language.} Paraphrasing reduces exposure of named people and avoids repeating
risky instructions, but prevents readers from checking the original wording.

\section{Conclusion}

Proving completion is not always a neutral afterthought. A listing may ask a
worker to reveal identity or location, use a personal account, act publicly or
physically, or stay available for later checks. This study offers an initial
vocabulary and manual audit for making those requests visible.

In this snapshot, severe proof burden---our label for scores of 4 or 5---is the majority case: \SevereSharePct\% of
eligible listings score severe, and they span
\SevereDistinctFeatureProfileCount\ combinations of requirements---so the
checklist, not a single score, names which requirements apply.

The share requesting at least one of physical-world action, location proof, or
recurring monitoring is higher among agent-or-bot-labeled listings in the full
sample and all three sensitivity analyses.
This is a hypothesis, not a confirmed difference: we created the
comparison after seeing the data; the \AgentRentAHumanListingCount\ agent-or-bot-labeled RentAHuman listings were posted under only 19 displayed names (20 including the Human Pages requester); category mixes differ; and requester-type labels remain unverified.
The measures are ready for worker testing, and the pattern is a hypothesis for a new, preplanned collection.

\section*{Data and Code Availability}

Only the supplementary PDF accompanies this paper. It contains additional
methods, aggregate results, and redacted examples, but excludes listing-level
data, raw task text, requester identifiers or hashes, follow-up records, and
coder or adjudicator packets. We invite researchers to contact us about collaboration or access to materials reduced to limit disclosure risk and reviewed before sharing. Any sharing would be considered case
by case and would remain subject to privacy, legal, institutional, licensing,
and platform-policy constraints.

\section*{Generative AI Disclosure}

Generative AI tools
helped revise this paper and supplement, review analyses, propose checks, locate
literature, edit code, and verify consistency and compilations.
The author reviewed every AI-assisted contribution retained in these documents,
made final decisions, and accepts full responsibility. The tools did not collect
listings, assign labels to listing content or study variables, or resolve coder
disagreements.

\bibliographystyle{ACM-Reference-Format}
\bibliography{references}


\begin{thebibliography}{48}


\ifx \showCODEN    \undefined \def \showCODEN     #1{\unskip}     \fi
\ifx \showDOI      \undefined \def \showDOI       #1{#1}\fi
\ifx \showISBNx    \undefined \def \showISBNx     #1{\unskip}     \fi
\ifx \showISBNxiii \undefined \def \showISBNxiii  #1{\unskip}     \fi
\ifx \showISSN     \undefined \def \showISSN      #1{\unskip}     \fi
\ifx \showLCCN     \undefined \def \showLCCN      #1{\unskip}     \fi
\ifx \shownote     \undefined \def \shownote      #1{#1}          \fi
\ifx \showarticletitle \undefined \def \showarticletitle #1{#1}   \fi
\ifx \showURL      \undefined \def \showURL       {\relax}        \fi
\providecommand\bibfield[2]{#2}
\providecommand\bibinfo[2]{#2}
\providecommand\natexlab[1]{#1}
\providecommand\showeprint[2][]{arXiv:#2}

\bibitem[Agapie et~al\mbox{.}(2015)]%
        {agapie2015field}
\bibfield{author}{\bibinfo{person}{Elena Agapie}, \bibinfo{person}{Jaime
  Teevan}, {and} \bibinfo{person}{Andr{\'e}s Monroy-Hern{\'a}ndez}.}
  \bibinfo{year}{2015}\natexlab{}.
\newblock \showarticletitle{Crowdsourcing in the Field: A Case Study Using
  Local Crowds for Event Reporting}.
\newblock \bibinfo{journal}{\emph{Proceedings of the AAAI Conference on Human
  Computation and Crowdsourcing}} \bibinfo{volume}{3}, \bibinfo{number}{1}
  (\bibinfo{year}{2015}), \bibinfo{pages}{2--11}.
\newblock
\urldef\tempurl%
\url{https://doi.org/10.1609/hcomp.v3i1.13235}
\showDOI{\tempurl}


\bibitem[Alkhatib et~al\mbox{.}(2017)]%
        {alkhatib2017piecework}
\bibfield{author}{\bibinfo{person}{Ali Alkhatib}, \bibinfo{person}{Michael~S.
  Bernstein}, {and} \bibinfo{person}{Margaret Levi}.}
  \bibinfo{year}{2017}\natexlab{}.
\newblock \showarticletitle{Examining Crowd Work and Gig Work Through The
  Historical Lens of Piecework}. In \bibinfo{booktitle}{\emph{Proceedings of
  the 2017 CHI Conference on Human Factors in Computing Systems}}
  \emph{(\bibinfo{series}{CHI '17})}. \bibinfo{publisher}{Association for
  Computing Machinery}, \bibinfo{address}{New York, NY, USA},
  \bibinfo{pages}{4599--4616}.
\newblock
\urldef\tempurl%
\url{https://doi.org/10.1145/3025453.3025974}
\showDOI{\tempurl}


\bibitem[Becher(2026)]%
        {becher2026rentahuman}
\bibfield{author}{\bibinfo{person}{Brooke Becher}.}
  \bibinfo{year}{2026}\natexlab{}.
\newblock \bibinfo{title}{At {RentAHuman}, {AI} Is the Boss and Humans Are
  `Meatworkers'}.
\newblock \bibinfo{howpublished}{Built In, March 11, 2026}.
\newblock
\urldef\tempurl%
\url{https://builtin.com/articles/what-is-rentahuman}
\showURL{%
\tempurl}


\bibitem[Breslow and Day(1980)]%
        {breslow1980casecontrol}
\bibfield{author}{\bibinfo{person}{Norman~E. Breslow} {and}
  \bibinfo{person}{Nicholas~E. Day}.} \bibinfo{year}{1980}\natexlab{}.
\newblock \bibinfo{booktitle}{\emph{Statistical Methods in Cancer Research,
  Volume I: The Analysis of Case-Control Studies}}.
\newblock Number~32 in \bibinfo{series}{IARC Scientific Publications}.
  \bibinfo{publisher}{International Agency for Research on Cancer},
  \bibinfo{address}{Lyon, France}.
\newblock
\showISBNx{978-92-832-0132-8}


\bibitem[Cameron et~al\mbox{.}(2008)]%
        {cameron2008bootstrap}
\bibfield{author}{\bibinfo{person}{A.~Colin Cameron}, \bibinfo{person}{Jonah~B.
  Gelbach}, {and} \bibinfo{person}{Douglas~L. Miller}.}
  \bibinfo{year}{2008}\natexlab{}.
\newblock \showarticletitle{Bootstrap-Based Improvements for Inference with
  Clustered Errors}.
\newblock \bibinfo{journal}{\emph{The Review of Economics and Statistics}}
  \bibinfo{volume}{90}, \bibinfo{number}{3} (\bibinfo{year}{2008}),
  \bibinfo{pages}{414--427}.
\newblock
\urldef\tempurl%
\url{https://doi.org/10.1162/rest.90.3.414}
\showDOI{\tempurl}


\bibitem[Cohen(1960)]%
        {cohen1960agreement}
\bibfield{author}{\bibinfo{person}{Jacob Cohen}.}
  \bibinfo{year}{1960}\natexlab{}.
\newblock \showarticletitle{A Coefficient of Agreement for Nominal Scales}.
\newblock \bibinfo{journal}{\emph{Educational and Psychological Measurement}}
  \bibinfo{volume}{20}, \bibinfo{number}{1} (\bibinfo{year}{1960}),
  \bibinfo{pages}{37--46}.
\newblock
\urldef\tempurl%
\url{https://doi.org/10.1177/001316446002000104}
\showDOI{\tempurl}


\bibitem[Cohen(1968)]%
        {cohen1968weighted}
\bibfield{author}{\bibinfo{person}{Jacob Cohen}.}
  \bibinfo{year}{1968}\natexlab{}.
\newblock \showarticletitle{Weighted Kappa: Nominal Scale Agreement with
  Provision for Scaled Disagreement or Partial Credit}.
\newblock \bibinfo{journal}{\emph{Psychological Bulletin}}
  \bibinfo{volume}{70}, \bibinfo{number}{4} (\bibinfo{year}{1968}),
  \bibinfo{pages}{213--220}.
\newblock
\urldef\tempurl%
\url{https://doi.org/10.1037/h0026256}
\showDOI{\tempurl}


\bibitem[Cronbach and Meehl(1955)]%
        {cronbach1955construct}
\bibfield{author}{\bibinfo{person}{Lee~J. Cronbach} {and}
  \bibinfo{person}{Paul~E. Meehl}.} \bibinfo{year}{1955}\natexlab{}.
\newblock \showarticletitle{Construct Validity in Psychological Tests}.
\newblock \bibinfo{journal}{\emph{Psychological Bulletin}}
  \bibinfo{volume}{52}, \bibinfo{number}{4} (\bibinfo{year}{1955}),
  \bibinfo{pages}{281--302}.
\newblock
\urldef\tempurl%
\url{https://doi.org/10.1037/h0040957}
\showDOI{\tempurl}


\bibitem[Diamantopoulos and Winklhofer(2001)]%
        {diamantopoulos2001formative}
\bibfield{author}{\bibinfo{person}{Adamantios Diamantopoulos} {and}
  \bibinfo{person}{Heidi~M. Winklhofer}.} \bibinfo{year}{2001}\natexlab{}.
\newblock \showarticletitle{Index Construction with Formative Indicators: An
  Alternative to Scale Development}.
\newblock \bibinfo{journal}{\emph{Journal of Marketing Research}}
  \bibinfo{volume}{38}, \bibinfo{number}{2} (\bibinfo{year}{2001}),
  \bibinfo{pages}{269--277}.
\newblock
\urldef\tempurl%
\url{https://doi.org/10.1509/jmkr.38.2.269.18845}
\showDOI{\tempurl}


\bibitem[D{\'i}az et~al\mbox{.}(2022)]%
        {diaz2022crowdworksheets}
\bibfield{author}{\bibinfo{person}{Mark D{\'i}az}, \bibinfo{person}{Ian~D.
  Kivlichan}, \bibinfo{person}{Rachel Rosen}, \bibinfo{person}{Dylan~K. Baker},
  \bibinfo{person}{Razvan Amironesei}, \bibinfo{person}{Vinodkumar
  Prabhakaran}, {and} \bibinfo{person}{Remi Denton}.}
  \bibinfo{year}{2022}\natexlab{}.
\newblock \showarticletitle{{CrowdWorkSheets}: Accounting for Individual and
  Collective Identities Underlying Crowdsourced Dataset Annotation}. In
  \bibinfo{booktitle}{\emph{Proceedings of the 2022 ACM Conference on Fairness,
  Accountability, and Transparency}} \emph{(\bibinfo{series}{FAccT '22})}.
  \bibinfo{publisher}{Association for Computing Machinery},
  \bibinfo{address}{New York, NY, USA}, \bibinfo{pages}{2342--2351}.
\newblock
\urldef\tempurl%
\url{https://doi.org/10.1145/3531146.3534647}
\showDOI{\tempurl}


\bibitem[Do et~al\mbox{.}(2024)]%
        {do2024sousveillance}
\bibfield{author}{\bibinfo{person}{Kimberly Do}, \bibinfo{person}{Maya
  De~Los~Santos}, \bibinfo{person}{Michael Muller}, {and}
  \bibinfo{person}{Saiph Savage}.} \bibinfo{year}{2024}\natexlab{}.
\newblock \showarticletitle{Designing Gig Worker Sousveillance Tools}. In
  \bibinfo{booktitle}{\emph{Proceedings of the CHI Conference on Human Factors
  in Computing Systems}} \emph{(\bibinfo{series}{CHI '24})}.
  \bibinfo{publisher}{Association for Computing Machinery},
  \bibinfo{address}{New York, NY, USA}, Article \bibinfo{articleno}{384},
  \bibinfo{numpages}{19}~pages.
\newblock
\urldef\tempurl%
\url{https://doi.org/10.1145/3613904.3642614}
\showDOI{\tempurl}


\bibitem[Gray and Suri(2019)]%
        {gray2019ghost}
\bibfield{author}{\bibinfo{person}{Mary~L. Gray} {and}
  \bibinfo{person}{Siddharth Suri}.} \bibinfo{year}{2019}\natexlab{}.
\newblock \bibinfo{booktitle}{\emph{Ghost Work: How to Stop Silicon Valley from
  Building a New Global Underclass}}.
\newblock \bibinfo{publisher}{Houghton Mifflin Harcourt},
  \bibinfo{address}{Boston, MA, USA}.
\newblock


\bibitem[Hara et~al\mbox{.}(2018)]%
        {hara2018earnings}
\bibfield{author}{\bibinfo{person}{Kotaro Hara}, \bibinfo{person}{Abi Adams},
  \bibinfo{person}{Kristy Milland}, \bibinfo{person}{Saiph Savage},
  \bibinfo{person}{Chris Callison-Burch}, {and} \bibinfo{person}{Jeffrey~P.
  Bigham}.} \bibinfo{year}{2018}\natexlab{}.
\newblock \showarticletitle{A Data-Driven Analysis of Workers' Earnings on
  Amazon Mechanical Turk}. In \bibinfo{booktitle}{\emph{Proceedings of the 2018
  CHI Conference on Human Factors in Computing Systems}}
  \emph{(\bibinfo{series}{CHI '18})}. \bibinfo{publisher}{Association for
  Computing Machinery}, \bibinfo{address}{New York, NY, USA}, Article
  \bibinfo{articleno}{449}, \bibinfo{numpages}{14}~pages.
\newblock
\urldef\tempurl%
\url{https://doi.org/10.1145/3173574.3174023}
\showDOI{\tempurl}


\bibitem[Hirth et~al\mbox{.}(2013)]%
        {hirth2013validation}
\bibfield{author}{\bibinfo{person}{Matthias Hirth}, \bibinfo{person}{Tobias
  Ho{\ss}feld}, {and} \bibinfo{person}{Phuoc Tran-Gia}.}
  \bibinfo{year}{2013}\natexlab{}.
\newblock \showarticletitle{Analyzing Costs and Accuracy of Validation
  Mechanisms for Crowdsourcing Platforms}.
\newblock \bibinfo{journal}{\emph{Mathematical and Computer Modelling}}
  \bibinfo{volume}{57}, \bibinfo{number}{11--12} (\bibinfo{year}{2013}),
  \bibinfo{pages}{2918--2932}.
\newblock
\urldef\tempurl%
\url{https://doi.org/10.1016/j.mcm.2012.01.006}
\showDOI{\tempurl}


\bibitem[Hu et~al\mbox{.}(2026)]%
        {hu2026boss}
\bibfield{author}{\bibinfo{person}{Qing Hu}, \bibinfo{person}{Qing Xiao},
  \bibinfo{person}{Hancheng Cao}, {and} \bibinfo{person}{Hong Shen}.}
  \bibinfo{year}{2026}\natexlab{}.
\newblock \showarticletitle{When Your Boss Is an {AI} Bot: Exploring
  Opportunities and Risks of Manager Clone Agents in the Future Workplace}. In
  \bibinfo{booktitle}{\emph{Proceedings of the 2026 CHI Conference on Human
  Factors in Computing Systems}} \emph{(\bibinfo{series}{CHI '26})}.
  \bibinfo{publisher}{Association for Computing Machinery},
  \bibinfo{address}{New York, NY, USA}, \bibinfo{pages}{1--21}.
\newblock
\urldef\tempurl%
\url{https://doi.org/10.1145/3772318.3790987}
\showDOI{\tempurl}


\bibitem[Ipeirotis et~al\mbox{.}(2010)]%
        {ipeirotis2010quality}
\bibfield{author}{\bibinfo{person}{Panagiotis~G. Ipeirotis},
  \bibinfo{person}{Foster Provost}, {and} \bibinfo{person}{Jing Wang}.}
  \bibinfo{year}{2010}\natexlab{}.
\newblock \showarticletitle{Quality Management on Amazon Mechanical Turk}. In
  \bibinfo{booktitle}{\emph{Proceedings of the ACM SIGKDD Workshop on Human
  Computation}} \emph{(\bibinfo{series}{HCOMP '10})}.
  \bibinfo{publisher}{Association for Computing Machinery},
  \bibinfo{address}{New York, NY, USA}, \bibinfo{pages}{64--67}.
\newblock
\urldef\tempurl%
\url{https://doi.org/10.1145/1837885.1837906}
\showDOI{\tempurl}


\bibitem[Irani and Silberman(2013)]%
        {irani2013turkopticon}
\bibfield{author}{\bibinfo{person}{Lilly~C. Irani} {and}
  \bibinfo{person}{M.~Six Silberman}.} \bibinfo{year}{2013}\natexlab{}.
\newblock \showarticletitle{Turkopticon: Interrupting Worker Invisibility in
  Amazon Mechanical Turk}. In \bibinfo{booktitle}{\emph{Proceedings of the
  SIGCHI Conference on Human Factors in Computing Systems}}
  \emph{(\bibinfo{series}{CHI '13})}. \bibinfo{publisher}{Association for
  Computing Machinery}, \bibinfo{address}{New York, NY, USA},
  \bibinfo{pages}{611--620}.
\newblock
\urldef\tempurl%
\url{https://doi.org/10.1145/2470654.2470742}
\showDOI{\tempurl}


\bibitem[Kamar(2016)]%
        {kamar2016hybrid}
\bibfield{author}{\bibinfo{person}{Ece Kamar}.}
  \bibinfo{year}{2016}\natexlab{}.
\newblock \showarticletitle{Directions in Hybrid Intelligence: Complementing
  {AI} Systems with Human Intelligence}. In
  \bibinfo{booktitle}{\emph{Proceedings of the Twenty-Fifth International Joint
  Conference on Artificial Intelligence}} \emph{(\bibinfo{series}{IJCAI '16})}.
  \bibinfo{publisher}{IJCAI/AAAI Press}, \bibinfo{pages}{4070--4073}.
\newblock


\bibitem[Kaur et~al\mbox{.}(2017)]%
        {kaur2017crowdmask}
\bibfield{author}{\bibinfo{person}{Harmanpreet Kaur}, \bibinfo{person}{Mitchell
  Gordon}, \bibinfo{person}{Yiwei Yang}, \bibinfo{person}{Jeffrey~P. Bigham},
  \bibinfo{person}{Jaime Teevan}, \bibinfo{person}{Ece Kamar}, {and}
  \bibinfo{person}{Walter~S. Lasecki}.} \bibinfo{year}{2017}\natexlab{}.
\newblock \showarticletitle{CrowdMask: Using Crowds to Preserve Privacy in
  Crowd-Powered Systems via Progressive Filtering}.
\newblock \bibinfo{journal}{\emph{Proceedings of the AAAI Conference on Human
  Computation and Crowdsourcing}} \bibinfo{volume}{5}, \bibinfo{number}{1}
  (\bibinfo{year}{2017}), \bibinfo{pages}{89--98}.
\newblock
\urldef\tempurl%
\url{https://doi.org/10.1609/hcomp.v5i1.13314}
\showDOI{\tempurl}


\bibitem[Kittur et~al\mbox{.}(2013)]%
        {kittur2013future}
\bibfield{author}{\bibinfo{person}{Aniket Kittur}, \bibinfo{person}{Jeffrey~V.
  Nickerson}, \bibinfo{person}{Michael Bernstein}, \bibinfo{person}{Elizabeth
  Gerber}, \bibinfo{person}{Aaron Shaw}, \bibinfo{person}{John Zimmerman},
  \bibinfo{person}{Matthew Lease}, {and} \bibinfo{person}{John Horton}.}
  \bibinfo{year}{2013}\natexlab{}.
\newblock \showarticletitle{The Future of Crowd Work}. In
  \bibinfo{booktitle}{\emph{Proceedings of the 2013 Conference on Computer
  Supported Cooperative Work}} \emph{(\bibinfo{series}{CSCW '13})}.
  \bibinfo{publisher}{Association for Computing Machinery},
  \bibinfo{address}{New York, NY, USA}, \bibinfo{pages}{1301--1318}.
\newblock
\urldef\tempurl%
\url{https://doi.org/10.1145/2441776.2441923}
\showDOI{\tempurl}


\bibitem[Kline and Santos(2012)]%
        {kline2012score}
\bibfield{author}{\bibinfo{person}{Patrick Kline} {and} \bibinfo{person}{Andres
  Santos}.} \bibinfo{year}{2012}\natexlab{}.
\newblock \showarticletitle{A Score Based Approach to Wild Bootstrap
  Inference}.
\newblock \bibinfo{journal}{\emph{Journal of Econometric Methods}}
  \bibinfo{volume}{1}, \bibinfo{number}{1} (\bibinfo{year}{2012}),
  \bibinfo{pages}{23--41}.
\newblock
\urldef\tempurl%
\url{https://doi.org/10.1515/2156-6674.1006}
\showDOI{\tempurl}


\bibitem[Lee(2026)]%
        {lee2026shadow}
\bibfield{author}{\bibinfo{person}{Lik-Hang Lee}.}
  \bibinfo{year}{2026}\natexlab{}.
\newblock \bibinfo{title}{The Shadow Boss: Identifying Atomized Manipulations
  in Agentic Employment of {XR} Users using Scenario Constructions}.
\newblock \bibinfo{howpublished}{arXiv preprint arXiv:2602.13622}.
\newblock
\urldef\tempurl%
\url{https://doi.org/10.48550/arXiv.2602.13622}
\showDOI{\tempurl}


\bibitem[Liang et~al\mbox{.}(2023)]%
        {liang2023monitoring}
\bibfield{author}{\bibinfo{person}{Chen Liang}, \bibinfo{person}{Jing Peng},
  \bibinfo{person}{Yili Hong}, {and} \bibinfo{person}{Bin Gu}.}
  \bibinfo{year}{2023}\natexlab{}.
\newblock \showarticletitle{The Hidden Costs and Benefits of Monitoring in the
  Gig Economy}.
\newblock \bibinfo{journal}{\emph{Information Systems Research}}
  \bibinfo{volume}{34}, \bibinfo{number}{1} (\bibinfo{year}{2023}),
  \bibinfo{pages}{297--318}.
\newblock
\urldef\tempurl%
\url{https://doi.org/10.1287/isre.2022.1130}
\showDOI{\tempurl}


\bibitem[MacKinnon et~al\mbox{.}(2025)]%
        {mackinnon2025logistic}
\bibfield{author}{\bibinfo{person}{James~G. MacKinnon},
  \bibinfo{person}{Morten~{\O}rregaard Nielsen}, {and}
  \bibinfo{person}{Matthew~D. Webb}.} \bibinfo{year}{2025}\natexlab{}.
\newblock \showarticletitle{Cluster-Robust Jackknife and Bootstrap Inference
  for Logistic Regression Models}.
\newblock \bibinfo{journal}{\emph{Econometric Reviews}} (\bibinfo{year}{2025}),
  \bibinfo{pages}{1--29}.
\newblock
\urldef\tempurl%
\url{https://doi.org/10.1080/07474938.2025.2515161}
\showDOI{\tempurl}


\bibitem[MacKinnon and Webb(2018)]%
        {mackinnon2018fewclusters}
\bibfield{author}{\bibinfo{person}{James~G. MacKinnon} {and}
  \bibinfo{person}{Matthew~D. Webb}.} \bibinfo{year}{2018}\natexlab{}.
\newblock \showarticletitle{The Wild Bootstrap for Few (Treated) Clusters}.
\newblock \bibinfo{journal}{\emph{The Econometrics Journal}}
  \bibinfo{volume}{21}, \bibinfo{number}{2} (\bibinfo{year}{2018}),
  \bibinfo{pages}{114--135}.
\newblock
\urldef\tempurl%
\url{https://doi.org/10.1111/ectj.12107}
\showDOI{\tempurl}


\bibitem[Mantel and Haenszel(1959)]%
        {mantel1959retrospective}
\bibfield{author}{\bibinfo{person}{Nathan Mantel} {and}
  \bibinfo{person}{William Haenszel}.} \bibinfo{year}{1959}\natexlab{}.
\newblock \showarticletitle{Statistical Aspects of the Analysis of Data from
  Retrospective Studies of Disease}.
\newblock \bibinfo{journal}{\emph{Journal of the National Cancer Institute}}
  \bibinfo{volume}{22}, \bibinfo{number}{4} (\bibinfo{year}{1959}),
  \bibinfo{pages}{719--748}.
\newblock
\urldef\tempurl%
\url{https://doi.org/10.1093/jnci/22.4.719}
\showDOI{\tempurl}


\bibitem[McInnis et~al\mbox{.}(2016)]%
        {mcinnis2016hit}
\bibfield{author}{\bibinfo{person}{Brian McInnis}, \bibinfo{person}{Dan
  Cosley}, \bibinfo{person}{Chaebong Nam}, {and} \bibinfo{person}{Gilly
  Leshed}.} \bibinfo{year}{2016}\natexlab{}.
\newblock \showarticletitle{Taking a HIT: Designing around Rejection, Mistrust,
  Risk, and Workers' Experiences in Amazon Mechanical Turk}. In
  \bibinfo{booktitle}{\emph{Proceedings of the 2016 CHI Conference on Human
  Factors in Computing Systems}} \emph{(\bibinfo{series}{CHI '16})}.
  \bibinfo{publisher}{Association for Computing Machinery},
  \bibinfo{address}{New York, NY, USA}, \bibinfo{pages}{2271--2282}.
\newblock
\urldef\tempurl%
\url{https://doi.org/10.1145/2858036.2858539}
\showDOI{\tempurl}


\bibitem[Mehta(2026)]%
        {mehta2026security}
\bibfield{author}{\bibinfo{person}{Pulak Mehta}.}
  \bibinfo{year}{2026}\natexlab{}.
\newblock \bibinfo{title}{Security Risks of {AI} Agents Hiring Humans: An
  Empirical Marketplace Study}.
\newblock \bibinfo{howpublished}{arXiv preprint arXiv:2602.19514}.
\newblock
\urldef\tempurl%
\url{https://doi.org/10.48550/arXiv.2602.19514}
\showDOI{\tempurl}


\bibitem[Moynihan et~al\mbox{.}(2015)]%
        {moynihan2015administrative}
\bibfield{author}{\bibinfo{person}{Donald Moynihan}, \bibinfo{person}{Pamela
  Herd}, {and} \bibinfo{person}{Hope Harvey}.} \bibinfo{year}{2015}\natexlab{}.
\newblock \showarticletitle{Administrative Burden: Learning, Psychological, and
  Compliance Costs in Citizen-State Interactions}.
\newblock \bibinfo{journal}{\emph{Journal of Public Administration Research and
  Theory}} \bibinfo{volume}{25}, \bibinfo{number}{1} (\bibinfo{year}{2015}),
  \bibinfo{pages}{43--69}.
\newblock
\urldef\tempurl%
\url{https://doi.org/10.1093/jopart/muu009}
\showDOI{\tempurl}


\bibitem[Nissenbaum(2004)]%
        {nissenbaum2004contextual}
\bibfield{author}{\bibinfo{person}{Helen Nissenbaum}.}
  \bibinfo{year}{2004}\natexlab{}.
\newblock \showarticletitle{Privacy as Contextual Integrity}.
\newblock \bibinfo{journal}{\emph{Washington Law Review}} \bibinfo{volume}{79},
  \bibinfo{number}{1} (\bibinfo{year}{2004}), \bibinfo{pages}{119--157}.
\newblock


\bibitem[Pradeep et~al\mbox{.}(2025)]%
        {pradeep2025gig}
\bibfield{author}{\bibinfo{person}{Amogh Pradeep}, \bibinfo{person}{Johanna
  Gunawan}, \bibinfo{person}{{\'A}lvaro Feal}, \bibinfo{person}{Woodrow
  Hartzog}, {and} \bibinfo{person}{David Choffnes}.}
  \bibinfo{year}{2025}\natexlab{}.
\newblock \showarticletitle{Gig Work at What Cost? Exploring Privacy Risks of
  Gig Work Platform Participation in the {U.S.}}
\newblock \bibinfo{journal}{\emph{Proceedings on Privacy Enhancing
  Technologies}} \bibinfo{volume}{2025}, \bibinfo{number}{1}
  (\bibinfo{year}{2025}), \bibinfo{pages}{491--510}.
\newblock
\urldef\tempurl%
\url{https://doi.org/10.56553/popets-2025-0027}
\showDOI{\tempurl}


\bibitem[Qian et~al\mbox{.}(2026a)]%
        {qian2026locating}
\bibfield{author}{\bibinfo{person}{Alice Qian}, \bibinfo{person}{Ryland Shaw},
  \bibinfo{person}{Laura Dabbish}, \bibinfo{person}{Jina Suh}, {and}
  \bibinfo{person}{Hong Shen}.} \bibinfo{year}{2026}\natexlab{a}.
\newblock \showarticletitle{Locating Risk: Task Designers and the Challenge of
  Risk Disclosure in Crowdsourced {RAI} Content Work}.
\newblock \bibinfo{journal}{\emph{Proceedings of the ACM on Human-Computer
  Interaction}} \bibinfo{volume}{10}, \bibinfo{number}{2}, Article
  \bibinfo{articleno}{CSCW029} (\bibinfo{year}{2026}),
  \bibinfo{numpages}{32}~pages.
\newblock
\urldef\tempurl%
\url{https://doi.org/10.1145/3788065}
\showDOI{\tempurl}


\bibitem[Qian et~al\mbox{.}(2026b)]%
        {qian2026discretion}
\bibfield{author}{\bibinfo{person}{Alice Qian}, \bibinfo{person}{Ziqi Yang},
  \bibinfo{person}{Ryland Shaw}, \bibinfo{person}{Jina Suh},
  \bibinfo{person}{Laura Dabbish}, {and} \bibinfo{person}{Hong Shen}.}
  \bibinfo{year}{2026}\natexlab{b}.
\newblock \showarticletitle{Worker Discretion Advised: Co-designing Risk
  Disclosure in Crowdsourced Responsible {AI} ({RAI}) Content Work}. In
  \bibinfo{booktitle}{\emph{Proceedings of the 2026 CHI Conference on Human
  Factors in Computing Systems}} \emph{(\bibinfo{series}{CHI '26})}.
  \bibinfo{publisher}{Association for Computing Machinery},
  \bibinfo{address}{New York, NY, USA}, \bibinfo{pages}{1--20}.
\newblock
\urldef\tempurl%
\url{https://doi.org/10.1145/3772318.3791558}
\showDOI{\tempurl}


\bibitem[Rosenblat and Stark(2016)]%
        {rosenblat2016algorithmic}
\bibfield{author}{\bibinfo{person}{Alex Rosenblat} {and} \bibinfo{person}{Luke
  Stark}.} \bibinfo{year}{2016}\natexlab{}.
\newblock \showarticletitle{Algorithmic Labor and Information Asymmetries: A
  Case Study of Uber's Drivers}.
\newblock \bibinfo{journal}{\emph{International Journal of Communication}}
  \bibinfo{volume}{10} (\bibinfo{year}{2016}), \bibinfo{pages}{3758--3784}.
\newblock


\bibitem[Salehi et~al\mbox{.}(2015)]%
        {salehi2015dynamo}
\bibfield{author}{\bibinfo{person}{Niloufar Salehi}, \bibinfo{person}{Lilly~C.
  Irani}, \bibinfo{person}{Michael~S. Bernstein}, \bibinfo{person}{Ali
  Alkhatib}, \bibinfo{person}{Eva Ogbe}, \bibinfo{person}{Kristy Milland},
  {and} \bibinfo{person}{Clickhappier}.} \bibinfo{year}{2015}\natexlab{}.
\newblock \showarticletitle{We Are Dynamo: Overcoming Stalling and Friction in
  Collective Action for Crowd Workers}. In
  \bibinfo{booktitle}{\emph{Proceedings of the 33rd Annual ACM Conference on
  Human Factors in Computing Systems}} \emph{(\bibinfo{series}{CHI '15})}.
  \bibinfo{publisher}{Association for Computing Machinery},
  \bibinfo{address}{New York, NY, USA}, \bibinfo{pages}{1621--1630}.
\newblock
\urldef\tempurl%
\url{https://doi.org/10.1145/2702123.2702508}
\showDOI{\tempurl}


\bibitem[Sannon and Cosley(2019)]%
        {sannon2019privacy}
\bibfield{author}{\bibinfo{person}{Shruti Sannon} {and} \bibinfo{person}{Dan
  Cosley}.} \bibinfo{year}{2019}\natexlab{}.
\newblock \showarticletitle{Privacy, Power, and Invisible Labor on Amazon
  Mechanical Turk}. In \bibinfo{booktitle}{\emph{Proceedings of the 2019 CHI
  Conference on Human Factors in Computing Systems}}
  \emph{(\bibinfo{series}{CHI '19})}. \bibinfo{publisher}{Association for
  Computing Machinery}, \bibinfo{address}{New York, NY, USA}, Article
  \bibinfo{articleno}{282}, \bibinfo{numpages}{12}~pages.
\newblock
\urldef\tempurl%
\url{https://doi.org/10.1145/3290605.3300512}
\showDOI{\tempurl}


\bibitem[Sannon et~al\mbox{.}(2022)]%
        {sannon2022gig}
\bibfield{author}{\bibinfo{person}{Shruti Sannon}, \bibinfo{person}{Billie
  Sun}, {and} \bibinfo{person}{Dan Cosley}.} \bibinfo{year}{2022}\natexlab{}.
\newblock \showarticletitle{Privacy, Surveillance, and Power in the Gig
  Economy}. In \bibinfo{booktitle}{\emph{Proceedings of the 2022 CHI Conference
  on Human Factors in Computing Systems}} \emph{(\bibinfo{series}{CHI '22})}.
  \bibinfo{publisher}{Association for Computing Machinery},
  \bibinfo{address}{New York, NY, USA}, Article \bibinfo{articleno}{619},
  \bibinfo{numpages}{15}~pages.
\newblock
\urldef\tempurl%
\url{https://doi.org/10.1145/3491102.3502083}
\showDOI{\tempurl}


\bibitem[Solove(2006)]%
        {solove2006taxonomy}
\bibfield{author}{\bibinfo{person}{Daniel~J. Solove}.}
  \bibinfo{year}{2006}\natexlab{}.
\newblock \showarticletitle{A Taxonomy of Privacy}.
\newblock \bibinfo{journal}{\emph{University of Pennsylvania Law Review}}
  \bibinfo{volume}{154}, \bibinfo{number}{3} (\bibinfo{year}{2006}),
  \bibinfo{pages}{477--560}.
\newblock


\bibitem[Tak(2026)]%
        {tak2026rented}
\bibfield{author}{\bibinfo{person}{Mudabbir~Ahmad Tak}.}
  \bibinfo{year}{2026}\natexlab{}.
\newblock \bibinfo{title}{Who Wants to be Rented? Rental Work and Digital
  Labour in the Age of {AI}}.
\newblock \bibinfo{howpublished}{SSRN 6560621}.
\newblock
\urldef\tempurl%
\url{https://doi.org/10.2139/ssrn.6560621}
\showDOI{\tempurl}


\bibitem[Tang et~al\mbox{.}(2026)]%
        {tang2026humantool}
\bibfield{author}{\bibinfo{person}{Yuanrong Tang}, \bibinfo{person}{Huiling
  Peng}, \bibinfo{person}{Bingxi Zhao}, \bibinfo{person}{Hengyang Ding},
  \bibinfo{person}{Hanchao Song}, \bibinfo{person}{Tianhong Wang},
  \bibinfo{person}{Chen Zhong}, {and} \bibinfo{person}{Jiangtao Gong}.}
  \bibinfo{year}{2026}\natexlab{}.
\newblock \bibinfo{title}{Human Tool: An {MCP}-Style Framework for Human-Agent
  Collaboration}.
\newblock \bibinfo{howpublished}{arXiv preprint arXiv:2602.12953}.
\newblock
\urldef\tempurl%
\url{https://doi.org/10.48550/arXiv.2602.12953}
\showDOI{\tempurl}


\bibitem[To et~al\mbox{.}(2014)]%
        {to2014location}
\bibfield{author}{\bibinfo{person}{Hien To}, \bibinfo{person}{Gabriel Ghinita},
  {and} \bibinfo{person}{Cyrus Shahabi}.} \bibinfo{year}{2014}\natexlab{}.
\newblock \showarticletitle{A Framework for Protecting Worker Location Privacy
  in Spatial Crowdsourcing}.
\newblock \bibinfo{journal}{\emph{Proceedings of the VLDB Endowment}}
  \bibinfo{volume}{7}, \bibinfo{number}{10} (\bibinfo{year}{2014}),
  \bibinfo{pages}{919--930}.
\newblock
\urldef\tempurl%
\url{https://doi.org/10.14778/2732951.2732966}
\showDOI{\tempurl}


\bibitem[Toxtli et~al\mbox{.}(2021)]%
        {toxtli2021invisible}
\bibfield{author}{\bibinfo{person}{Carlos Toxtli}, \bibinfo{person}{Siddharth
  Suri}, {and} \bibinfo{person}{Saiph Savage}.}
  \bibinfo{year}{2021}\natexlab{}.
\newblock \showarticletitle{Quantifying the Invisible Labor in Crowd Work}.
\newblock \bibinfo{journal}{\emph{Proceedings of the ACM on Human-Computer
  Interaction}} \bibinfo{volume}{5}, \bibinfo{number}{CSCW2}, Article
  \bibinfo{articleno}{319} (\bibinfo{year}{2021}),
  \bibinfo{numpages}{26}~pages.
\newblock
\urldef\tempurl%
\url{https://doi.org/10.1145/3476060}
\showDOI{\tempurl}


\bibitem[Vallas and Schor(2020)]%
        {vallas2020platforms}
\bibfield{author}{\bibinfo{person}{Steven Vallas} {and}
  \bibinfo{person}{Juliet~B. Schor}.} \bibinfo{year}{2020}\natexlab{}.
\newblock \showarticletitle{What Do Platforms Do? Understanding the Gig
  Economy}.
\newblock \bibinfo{journal}{\emph{Annual Review of Sociology}}
  \bibinfo{volume}{46} (\bibinfo{year}{2020}), \bibinfo{pages}{273--294}.
\newblock
\urldef\tempurl%
\url{https://doi.org/10.1146/annurev-soc-121919-054857}
\showDOI{\tempurl}


\bibitem[van Zoonen et~al\mbox{.}(2026)]%
        {vanzoonen2026algorithmic}
\bibfield{author}{\bibinfo{person}{Ward van Zoonen}, \bibinfo{person}{Monika~E.
  von Bonsdorff}, {and} \bibinfo{person}{Beatrice I. J.~M. van~der Heijden}.}
  \bibinfo{year}{2026}\natexlab{}.
\newblock \showarticletitle{Algorithmic Surveillance and Workers' Compliance:
  The Role of Trust, Privacy Concerns, and Fairness in Online Crowdwork}.
\newblock \bibinfo{journal}{\emph{Human Relations}} \bibinfo{volume}{79},
  \bibinfo{number}{7} (\bibinfo{year}{2026}), \bibinfo{pages}{795--824}.
\newblock
\urldef\tempurl%
\url{https://doi.org/10.1177/00187267251379698}
\showDOI{\tempurl}


\bibitem[Wilkins(2026)]%
        {wilkins2026rent}
\bibfield{author}{\bibinfo{person}{Joe Wilkins}.}
  \bibinfo{year}{2026}\natexlab{}.
\newblock \bibinfo{title}{New Site Lets {AI} Rent Human Bodies}.
\newblock \bibinfo{howpublished}{Futurism, February 4, 2026}.
\newblock
\urldef\tempurl%
\url{https://futurism.com/artificial-intelligence/ai-rent-human-bodies}
\showURL{%
\tempurl}


\bibitem[Wilson(1927)]%
        {wilson1927probable}
\bibfield{author}{\bibinfo{person}{Edwin~B. Wilson}.}
  \bibinfo{year}{1927}\natexlab{}.
\newblock \showarticletitle{Probable Inference, the Law of Succession, and
  Statistical Inference}.
\newblock \bibinfo{journal}{\emph{J. Amer. Statist. Assoc.}}
  \bibinfo{volume}{22}, \bibinfo{number}{158} (\bibinfo{year}{1927}),
  \bibinfo{pages}{209--212}.
\newblock
\urldef\tempurl%
\url{https://doi.org/10.1080/01621459.1927.10502953}
\showDOI{\tempurl}


\bibitem[Xia et~al\mbox{.}(2017)]%
        {xia2017privacy}
\bibfield{author}{\bibinfo{person}{Huichuan Xia}, \bibinfo{person}{Yang Wang},
  \bibinfo{person}{Yun Huang}, {and} \bibinfo{person}{Anuj Shah}.}
  \bibinfo{year}{2017}\natexlab{}.
\newblock \showarticletitle{``Our Privacy Needs to Be Protected at All Costs'':
  Crowd Workers' Privacy Experiences on Amazon Mechanical Turk}.
\newblock \bibinfo{journal}{\emph{Proceedings of the ACM on Human-Computer
  Interaction}} \bibinfo{volume}{1}, \bibinfo{number}{CSCW}, Article
  \bibinfo{articleno}{113} (\bibinfo{year}{2017}),
  \bibinfo{numpages}{22}~pages.
\newblock
\urldef\tempurl%
\url{https://doi.org/10.1145/3134748}
\showDOI{\tempurl}


\bibitem[Zuboff(2019)]%
        {zuboff2019surveillance}
\bibfield{author}{\bibinfo{person}{Shoshana Zuboff}.}
  \bibinfo{year}{2019}\natexlab{}.
\newblock \bibinfo{booktitle}{\emph{The Age of Surveillance Capitalism: The
  Fight for a Human Future at the New Frontier of Power}}.
\newblock \bibinfo{publisher}{PublicAffairs}, \bibinfo{address}{New York, NY,
  USA}.
\newblock


\end{thebibliography}

\end{document}